\documentclass[preprint,pre,aps,amsfonts,amsmath,showpacs,superscriptaddress,nofootinbib, draft]{revtex4}
\usepackage{bm}
\begin{document}
\title{Wave Localization Does not Affect the Breakdown of a Schr\"odinger-Type Amplifier Driven by the Square of a Gaussian Field}
\author{Philippe Mounaix}
\email{mounaix@cpht.polytechnique.fr}
\author{Pierre Collet}
\email{collet@cpht.polytechnique.fr}
\affiliation{Centre de Physique Th\'eorique, UMR 7644 du CNRS, Ecole
Polytechnique, 91128 Palaiseau Cedex, France.}
\date{\today}
\begin{abstract}
We study the divergence of the solution to a Schr\"odinger-type amplifier driven by the square of a Gaussian noise in presence of a random potential.
We follow the same approach as Mounaix, Collet, and Lebowitz (MCL) in terms of a distributional formulation of the amplified field and the use of the Paley-Wiener theorem [Commun. Math. Phys. {\bf 264}, 741-758 (2006) and {\bf 280}, 281-283 (2008)]. Our results show that the divergence is not affected by the random potential, in the sense that it occurs at exactly the same coupling constant as what was found by MCL without a potential. It follows {\it a fortiori} that the breakdown of the amplifier is not affected by the possible existence of a localized regime in the amplification free limit.
\end{abstract}
\pacs{42.25.Dd, 02.50.Ey, 52.38.-r, 72.15.Rn}
\maketitle
%
%
\newtheorem{lemma}{Lemma}
\newtheorem{proposition}{Proposition}
\newtheorem{corollary}{Corollary}
\section{Introduction}\label{sec1}
We investigate the breakdown of linear amplification in a system driven by the square of a Gaussian noise in presence of a random potential. We consider the stochastic PDE
\begin{equation}\label{eq1.1}
\left\lbrace\begin{array}{l}
\partial_t{\cal E}(x,t)-\frac{i}{2m}\Delta
{\cal E}(x,t)=
\left\lbrack\lambda\vert S(x,t)\vert^2-i\rho(x,t)\right\rbrack{\cal E}(x,t),\\
t\ge 0,\ x\in\Lambda\subset {\mathbb R}^d,\ {\rm and}\ {\cal E}(x,0)=1,
\end{array}\right.
\end{equation}
where $m\ne 0$ is a complex mass with ${\rm Im}(m)\ge 0$, $\lambda >0$ is the coupling constant, $S$ is a zero mean complex Gaussian noise, and $\rho$ is a zero mean real noise (not necessarily Gaussian). In the ``diffractive case" where ${\rm Im}(m)=0$ and ${\rm Re}(m)\ne 0$, this problem models the backscattering of an incoherent laser by an optically active medium with a randomly perturbed index of refraction\ \cite{note1}.

In the unperturbed limit $\rho(x,t)\equiv 0$, the breakdown of\ (\ref{eq1.1}) defined as the divergence of its average solution was investigated in\ \cite{MCL}. There, the value of $\lambda$ at which the $q$-th moment of $\vert {\cal E}(x,t)\vert$ w.r.t. $S$ diverges is obtained. Quite remarkably, this value is found to be independent of $m$ for $\vert m\vert^{-1}>0$ and always less or equal for $\vert m\vert^{-1}>0$ than for $\vert m\vert^{-1}=0$, i.e. when the $\Delta {\cal E}$ term is absent. This somewhat surprising result follows from the fact that, however small $\vert m\vert^{-1}>0$ is, the $\Delta {\cal E}$ term allows amplification to sample every ray trajectory. In particular, the most amplified paths always contribute to the amplification of ${\cal E}$ and it can be shown that the breakdown of\ (\ref{eq1.1}) results from the divergence of their contributions. On the contrary, if $\vert m\vert^{-1}=0$ only amplification along straight paths can contribute to the overall amplification of ${\cal E}$ and cause its divergence.

Equation\ (\ref{eq1.1}) with a time independent potential has been extensively studied in the opposite case of no amplification, i.e. with $\rho(x,t)\equiv\rho(x)$ and $\lambda =0$\ \cite{Tho}-\cite{Joh}. For $d\le 2$, the situation of interest in optics\ \cite{note1}, the random Hamiltonian $H_0=-(2m)^{-1}\Delta +\rho(x)$ has only point spectrum with localized eigenfunctions randomly distributed over $\Lambda$. When amplification is turned on ($\lambda >0$), the most amplified paths may happen to zigzag across different eigenfunctions of $H_0$. In this case, the contribution of the most amplified paths is expected to be reduced relatively to the one of less amplified, but straighter, ray trajectories bound to only one localized eigenfunction. The question then arises whether such localization effects are powerful enough to make the breakdown of\ (\ref{eq1.1}) occur at a greater $\lambda$ than for $\rho(x,t)\equiv 0$.

In this paper we answer that question by determining the value of $\lambda$ at which the $q$-th moment of $\vert {\cal E}(x,t)\vert$ w.r.t. $S$ diverges for almost every realization of $\rho$. We follow the same strategy as in\ \cite{MCL}. Considering a wide class of time dependent $\rho(x,t)$, we find that this value  is the same as when $\rho(x,t)\equiv 0$ (the result holds for almost every realization of $\rho$). The possible existence of a localized regime for the solution to\ (\ref{eq1.1}) with $\lambda =0$ does not affect the divergence of its moments when $\lambda >0$.

The outline of the paper is as follows. In Sect.\ \ref{sec2} we specify the classes of $\rho$ and $S$ which we can treat and we give some definitions. Sect.\ \ref{sec3} is devoted to technical results yielding the control of the growth of $\vert {\cal E}(x,t)\vert^q$. The divergence of the moments of $\vert {\cal E}(x,t)\vert$ w.r.t. $S$ for almost every realization of $\rho$ is investigated in Sect.\ \ref{sec4}. Finally, the divergence of the moments of $\vert {\cal E}(x,t)\vert$ w.r.t. both $\rho$ and $S$ is investigated for a slightly reduced class of $\rho$ at the end of Sect.\ \ref{sec4}.
%
%
\section{Model and definitions}\label{sec2}
We consider the solution to the linear amplifier equation\ (\ref{eq1.1}) with $m$ in $\overline{{\mathbb C}^+}\backslash\lbrace 0\rbrace$, where $\overline{{\mathbb C}^+} \equiv\lbrace m\in {\mathbb C}:\ {\rm Im}(m)\ge 0\rbrace$, and $\Lambda$ a $d$-dimensional torus with $d\le 3$. The random field $S$ is the same as in\ \cite{MCL}. Namely, we assume that $S$ can be expressed as a finite combination of $M$ complex Gaussian r.v., $s_n$,
\begin{equation}\label{eq2.1}
S(x,t)=\sum_{n=1}^{M}s_n\Phi_n(x,t),
\end{equation}
with
\begin{equation}\label{eq2.2}
\left\lbrace\begin{array}{l}
\langle s_n\rangle =\langle s_n s_m\rangle =0,\\
\langle s_n s_m^\ast\rangle =\delta_{nm}.
\end{array}\right.
\end{equation}
The $\Phi_n$ are normalized such that
$$
\frac{1}{\vert\Lambda\vert}\int_0^1\int_\Lambda \langle\vert S(x,\tau)\vert^2\rangle\, d\tau d^dx =
\frac{1}{\vert\Lambda\vert}\sum_{n=1}^M\int_0^1\int_\Lambda \vert\Phi_n(x,\tau)\vert^2\, d\tau d^dx =
1.
$$
Furthermore, the $\Phi_n(\cdot ,\tau)$ are assumed to have second derivatives bounded uniformly in $\tau\in [0,t]$, and the $\Phi_n(x,\cdot)$ are piecewise continuous for every $x\in\Lambda$ with a finite number of discontinuities in $[0,t]$ for all finite $t$.

Let $\|\cdot\|_\infty$ denote the uniform norm on $\Lambda$ and write $\vert\vert\vert\cdot\vert\vert\vert_\infty =\sup_{0\le\tau\le t}\|\cdot\|_\infty$. The random potential $\rho(x,t)$ is assumed to be independent of $S(x,t)$ and such that, with probability one, $\vert\vert\vert\rho\vert\vert\vert_\infty <+\infty$, $\vert\vert\vert\nabla\rho\vert\vert\vert_\infty <+\infty$, and $\vert\vert\vert\Delta\rho\vert\vert\vert_\infty <+\infty$.

Let $\langle\cdot\rangle_s$ and $\langle\cdot\rangle_\rho$ denote the statistical averages w.r.t. $S$ and $\rho$, respectively.  We are interested in the critical couplings $\lambda_{q}^{(-)}(x,t)$, $\lambda_{q}^{(+)}(x,t)$, and $\lambda_{q}(x,t)$ defined by
\begin{subequations}\label{eq2.3}
\begin{eqnarray}
&&\lambda_{q}^{(-)}(x,t)=
\sup\lbrace \lambda\ge 0\ {\rm below\ which}\ 
\langle\vert {\cal E}(x,t)\vert^q\rangle_s <+\infty\ \ {\rm a.\, s.}\rbrace,
\label{eq2.3a}\\
&&\lambda_{q}^{(+)}(x,t)=
\inf\lbrace \lambda> 0\ {\rm above\ which}\ 
\langle\vert {\cal E}(x,t)\vert^q\rangle_s =+\infty\ \ {\rm a.\, s.}\rbrace,
\label{eq2.3b}\\
&&\lambda_{q}(x,t)=
\inf\lbrace \lambda>0:
\langle\vert {\cal E}(x,t)\vert^q\rangle_{\rho ,s} =+\infty\rbrace.
\label{eq2.3c}
\end{eqnarray}
\end{subequations}
In\ (\ref{eq2.3a}) and\ (\ref{eq2.3b}) ``almost surely" refers to the appropriate probability measure for $\rho$. Note that neither $S$ nor $\rho$ are assumed to be homogeneous and the critical coupling will depend on $x$ in general.
%
%
\section{Controlling the growth of $\bm{\vert{\cal E}(x,t)\vert^q}$}\label{sec3}
In this section we go back over the sections III and IV of\ \cite{MCL} allowing for the presence of the $\rho(x,t)$ random term on the right-hand side of\ (\ref{eq1.1}). Since the calculations are essentially the same, we do not give all the intermediate steps and refer the interested reader to\ \cite{MCL} for details.

Let $s$ be the $M$-dimensional Gaussian random vector the elements of which are the $s_n$, and $\gamma (x,\tau)$ the $M\times M$ Hermitian matrix defined by
$$
\gamma_{nm}(x,\tau)=\Phi_n^\ast(x,\tau)\Phi_m(x,\tau).
$$
Let $\varphi_i$ be $N$ real valued functions, with $N=M^2$, given by $\gamma_{nn}$, $\sqrt{2}{\rm Re}(\gamma_{nm})$, and $\sqrt{2}{\rm Im}(\gamma_{nm})$, $n<m$. Define
\begin{equation}\label{eq3.1}
G_{x,t;\rho}(u)=\frac{1}{(2\pi)^N}\int\cdots\int_{{\mathbb R}^N} \Psi(x,t;\eta ,\rho)\, {\rm e}^{iu\cdot\eta}
\prod_{i=1}^N d\eta_i ,
\end{equation}
in which $\Psi(x,t;\eta ,\rho)$ is the solution to the Schr\"odinger equation
\begin{equation}\label{eq3.2}
\left\lbrace\begin{array}{l}
i\partial_t\Psi(x,t;\eta, \rho)=-\frac{1}{2m}\Delta\Psi(x,t;\eta ,\rho)
+\lbrack\rho(x,t)+V(x,t;\eta)\rbrack\Psi(x,t;\eta ,\rho),\\
t\ge 0,\ x\in\Lambda,\ {\rm and}\ \Psi(x,0;\eta ,\rho)=1,
\end{array}\right. 
\end{equation}
where $V(x,t;\eta)$ is given by
\begin{equation}\label{eq3.3}
V(x,t;\eta)\equiv\sum_{i=1}^N\eta_i\varphi_i(x,t).
\end{equation}
In the following we make the dependence of ${\cal E}(x,t)$ on the realizations of $s$ and $\rho$ explicit by writing ${\cal E}(x,t)\equiv {\cal E}(x,t;s,\rho)$.  Let $a_i=\inf_{x(\cdot)\in B(x,t)}\int_0^t\varphi_i(x(\tau),\tau)\, d\tau$ and $b_i=\sup_{x(\cdot)\in B(x,t)}\int_0^t\varphi_i(x(\tau),\tau)\, d\tau$, where $B(x,t)$ denotes the set of all the continuous paths in $\Lambda$ satisfying $x(t)=x$. Let $k(s)$ be a vector in $\mathbb{R}^N$ the components of which are given by $\vert s_n\vert^2$, $\sqrt{2}{\rm Re}(s_n s_m^\ast)$, and $\sqrt{2}{\rm Im}(s_n s_m^\ast)$, $n<m$. One has,
\begin{lemma}\label{lem1}
For every $t>0$, $x\in\Lambda$, and $m\in\overline{{\mathbb C}^+}\backslash\lbrace 0\rbrace$,
\begin{enumerate}
\item[(i)] for almost every realization of $\rho$, $G_{x,t;\rho}$ defined by\ (\ref{eq3.1}) is a distribution with compact support on ${\mathbb R}^N$ and ${\rm supp}G_{x,t}\subset\lbrack a_1,b_1\rbrack\times ...\times\lbrack a_N,b_N\rbrack$;
\item[(ii)] for almost every realization of $\rho$ the solution to\ (\ref{eq1.1}) is given by
\end{enumerate}
\begin{equation}\label{eq3.4}
{\cal E}(x,t;s,\rho)=\int\cdots\int_{{\mathbb R}^N}
G_{x,t;\rho}(u)\, {\rm e}^{\lambda k(s)\cdot u}\prod_{i=1}^N du_i.
\end{equation}
\end{lemma}
{\it Proof.} Analyticity of $\Psi(x,t;\eta ,\rho)$ in $\eta$ is proved in Appendix (see the proof of Lemma\ \ref{lem3})\ \cite{note2}. Let $\tilde{\Psi}(x,t;\eta ,\rho)=\Psi(x,t;\eta ,\rho)\exp(-it\sum_{i=1}^N\eta_ic_i)$ where the constants $c_i\in {\mathbb R}$ are given by
\begin{equation}\label{eq3.5}
c_i=-\frac{1}{2t}\left\lbrace
\int_0^t\sup_{x\in\Lambda}\lbrack\varphi_i(x,\tau)\rbrack\, d\tau
+\int_0^t\inf_{x\in\Lambda}\lbrack\varphi_i(x,\tau)\rbrack\, d\tau
\right\rbrace .
\end{equation}
$\tilde{\Psi}$ is the solution to\ (\ref{eq3.2}) with $V$ given by\ (\ref{eq3.3}) in which the $\varphi_i$ are replaced with $\tilde{\varphi}_i=\varphi_i+c_i$. Let $\epsilon_i={\rm sgn}\lbrack {\rm Im}(\eta_i)\rbrack$ and define
\begin{equation}\label{eq3.6}
\kappa_i\equiv\int_0^t\sup_{x\in\Lambda}
\lbrack\epsilon_i\tilde{\varphi}_i(x,\tau)\rbrack\, d\tau
=\frac{1}{2}\left\lbrace
\int_0^t\sup_{x\in\Lambda}\lbrack\varphi_i(x,\tau)\rbrack\, d\tau
-\int_0^t\inf_{x\in\Lambda}\lbrack\varphi_i(x,\tau)\rbrack\, d\tau
\right\rbrace .
\end{equation}
Note that, with this choice of $c_i$,  $\kappa_i$ is independent of $\epsilon_i$. Making the same calculation as in\ \cite{MCL} [from Equation (15) to Equation (21)] with the random potential $\rho(x,t)$ on the right-hand side of\ (\ref{eq3.2}), one finds that
\begin{eqnarray}\label{eq3.7}
\vert\tilde{\Psi}(x,t;\eta ,\rho)\vert\le&&\left\lbrack A+Bt\left(\vert\vert\vert\Delta\rho\vert\vert\vert_\infty +C\sum_{i=1}^N\vert\eta_i\vert\right)
+Bt^2\left(\vert\vert\vert\nabla\rho\vert\vert\vert_\infty +D\sum_{i=1}^N\vert\eta_i\vert\right)^2\right\rbrack \nonumber \\
&\times&{\rm e}^{\sum_{i=1}^N \kappa_i\vert\eta_i\vert},
\end{eqnarray}
where $A$, $B$, $C$, and $D$ are finite and independent of $\eta$ and $m$. Since both $\vert\vert\vert\nabla\rho\vert\vert\vert_\infty$ and $\vert\vert\vert\Delta\rho\vert\vert\vert_\infty$ are almost surely bounded by assumption,\ (\ref{eq3.7}) is similar to the equation (21) in\ \cite{MCL}, with probability one. The same reasoning as in the paragraph below the equation (21) in\ \cite{MCL} completes the proof of {\it (i)}.

To prove {\it (ii)} it suffices to note that according to\ (\ref{eq3.1}),\ (\ref{eq3.4}) can be rewritten as ${\cal E}(x,t;s ,\rho)=\Psi(x,t;\eta =i\lambda k(s),\rho)$ which is the solution to\ (\ref{eq3.2}) with $\eta =i\lambda k(s)$ in the potential\ (\ref{eq3.3}). It can be checked that the latter equation is indeed Equation\ (\ref{eq1.1}) [reconstruct $\vert S\vert^2=s^\dag\gamma s$ from its monomial decomposition and multiply\ (\ref{eq3.3}) by $-i$], which completes the proof of Lemma 1. $\square$

As explained in\ \cite{MCL}, Lemma\ \ref{lem1} makes it possible to use the Paley-Wiener theorem in order to control the growth of ${\cal E}(x,t;s)$ as $\| s\|\rightarrow +\infty$. Let $\hat{s}\equiv s/\| s\|$ be the direction of $s$ in ${\mathbb C}^M$ and $H_{x,t}(\hat{s})=\sup_{x(\cdot)\in B(x,t)}\int_0^t U(x(\tau),\tau;\hat{s})\, d\tau$, with $U(x,\tau;\hat{s})=\sum_{i=1}^N\hat{k}(s)_i\varphi_i(x,\tau)$ where $\hat{k}(s)= k(s)/\| k(s)\|$. One has the following lemma,
\begin{lemma}\label{lem2}
For every $t>0$, $x\in\Lambda$, $m\in\overline{{\mathbb C}^+}\backslash\lbrace 0\rbrace$, and $q$ a positive integer, one has
\begin{equation}\label{eq3.8}
\limsup_{\| s\|\rightarrow +\infty}\frac{\ln\left\vert {\cal E}(x,t;s,\rho)\right\vert^q}
{\| s\|^2}= q\lambda H_{x,t}(\hat{s}),
\end{equation}
along every given direction $\hat{s}$ in ${\mathbb C}^M$ and for almost every realization of $\rho$.
\end{lemma}
{\it Proof.} According to Lemma\ \ref{lem1}, one is allowed to follow the same line as the one leading to the equation (24) in\ \cite{MCL} for almost every realization of $\rho$. Thus,
\begin{equation}\label{eq3.9}
\limsup_{\| s\|\rightarrow +\infty}\frac{\ln\left\vert {\cal E}(x,t;s,\rho)\right\vert^q}
{\| s\|^2}\le  q\lambda H_{x,t}(\hat{s}),
\end{equation}
along every direction $\hat{s}$ in ${\mathbb C}^M$ and for almost every realization of $\rho$. Let $g_{x,t;\rho}(u)$ be a distribution with compact support on ${\mathbb R}$ whose Fourier transform, $\Psi^{(m)}(x,t;\eta ,\rho)\equiv({\cal F}g_{x,t;\rho}^{(m)})(\eta)$ with $\eta\in {\mathbb R}$, is the solution to\ (\ref{eq3.2}) with $V(x,t;\eta)=\eta U(x,t;\hat{s})$. Inequality\ (\ref{eq3.9}) reduces to an equality if one can prove that $\sup\lbrace v\in\mathbb{R}:v\in {\rm supp}g_{x,t;\rho}\rbrace =H_{x,t}(\hat{s})$ for almost every realization of $\rho$ (see the end of the proof of Lemma 2 in\ \cite{MCL}). This is done in Appendix. $\square$
%
%
\section{Determination of the critical couplings}\label{sec4}
In this section we prove that the presence of the random potential $\rho(x,t)$ on the right-hand side of\ (\ref{eq1.1}) does not affect the value of the critical coupling obtained in\ \cite{MCL}.

Let $\mu_1\lbrack x(\cdot)\rbrack>0$ be the largest eigenvalue of the covariance operator $T_{x(\cdot)}$ acting on
$f(\tau)\in L^2(d\tau)$, defined by
$$
(T_{x(\cdot)} f)(\tau)=\int_0^t \langle S^\ast (x(\tau),\tau)S(x(\tau'),\tau')\rangle f(\tau')\, d\tau',
$$
with $0\le\tau,\tau'\le t$ and $x(\cdot)\in B(x,t)$. It is shown in\ \cite{MCL} that there is a one-to-one relationship between the non vanishing eigenvalues of $T_{x(\cdot)}$ and those of the matrix $\int_0^t\gamma(x(\tau),\tau)\, d\tau$. In particular, $\mu_1\lbrack x(\cdot)\rbrack$ is also the largest eigenvalue of $\int_0^t\gamma(x(\tau),\tau)\, d\tau$. Define
\begin{equation}\label{eq4.1}
\mu_{x,t}=\sup_{x(\cdot)\in B(x,t)}\mu_1\lbrack x(\cdot)\rbrack .
\end{equation}
The critical couplings $\lambda_{q}^{(-)}(x,t)$ and $\lambda_{q}^{(-)}(x,t)$ defined in\ (\ref{eq2.3a}) and\ (\ref{eq2.3b}) are given by the following proposition.
\begin{proposition}\label{prop1}
For every $t>0$ and $x\in\Lambda$, $\lambda_{q}^{(-)}(x,t)=\lambda_{q}^{(+)}(x,t)=(q\mu_{x,t})^{-1}$.
\end{proposition}
{\it Proof.} First we prove $\lambda_{q}^{(-)}(x,t)\ge (q\mu_{x,t})^{-1}$. Expressing $U(x(\tau),\tau;\hat{s})$ in terms of the quadratic form $s^{\dag}\gamma (x(\tau),\tau)s$ in the expression for $H_{x,t}(\hat{s})$ (see the paragraph above Lemma\ \ref{lem2}), one has
$$
H_{x,t}(\hat{s})=\sup_{x(\cdot)\in B(x,t)}\, \frac{s^\dag}{\vert\vert s\vert\vert}
\left\lbrack\int_0^t\gamma(x(\tau),\tau)\, d\tau\right\rbrack
\frac{s}{\vert\vert s\vert\vert} \le
\mu_{x,t} .
$$
Hence, by Lemma 2,
$$
\limsup_{\| s\|\rightarrow +\infty}\frac{\ln\left\vert {\cal E}(x,t;s,\rho)\right\vert^q}
{\| s\|^2}\le q\lambda\mu_{x,t} ,
$$
for almost every realization of $\rho$. This implies that for every $\lambda <(q\mu_{x,t})^{-1}$ and almost every realization of $\rho$,
\begin{equation}\label{eq4.2}
\langle\vert {\cal E}(x,t)\vert^q\rangle_s =
\int\cdots\int_{{\mathbb C}^M} {\rm e}^{-\| s\|^2}
\vert {\cal E}(x,t;s,\rho)\vert^q\prod_{n=1}^M \frac{d^2s_n}{\pi}<+\infty ,
\end{equation}
which proves $\lambda_{q}^{(-)}(x,t)\ge (q\mu_{x,t})^{-1}$.

We now prove $\lambda_{q}^{(+)}(x,t)\le (q\mu_{x,t})^{-1}$. According to Lemmas\ \ref{lem1} and\ \ref{lem2}, the calculation in the second Ref.\ \cite{MCL} can be carried out for almost every realisation of $\rho$, yielding $\langle\vert {\cal E}(x,t)\vert^q\rangle_s =+\infty$ for every $\lambda >(q\mu_{x,t})^{-1}$ and almost every realisation of $\rho$. Therefore $\lambda_{q}^{(+)}(x,t)\le (q\mu_{x,t})^{-1}$. The obvious inequality $\lambda_{q}^{(-)}(x,t)\le\lambda_{q}^{(+)}(x,t)$ completes the proof. $\square$

As a corollary of Proposition\ \ref{prop1}, one gets the following upper bound for the third critical coupling $\lambda_{q}(x,t)$, defined in\ (\ref{eq2.3c}).
\begin{corollary}\label{coro1}
For every $t>0$ and $x\in\Lambda$, $\lambda_{q}(x,t)\le(q\mu_{x,t})^{-1}$.
\end{corollary}
{\it Proof.} By proposition\ \ref{prop1}, $\langle\vert {\cal E}(x,t)\vert^q\rangle_s =+\infty$ for every $\lambda >(q\mu_{x,t})^{-1}$ and almost every realisation of $\rho$. Thus $\langle\vert {\cal E}(x,t)\vert^q\rangle_{\rho ,s} =\langle\vert {\cal E}(x,t)\vert^q\rangle_{s,\rho} =+\infty$ for every $\lambda >(q\mu_{x,t})^{-1}$, which proves the corollary. $\square$

Considering a slightly reduced class of $\rho$, it is possible to go beyond Corollary\ \ref{coro1} and establish the counterpart of Proposition\ \ref{prop1} for $\lambda_{q}(x,t)$. This is the subject of the following proposition.
\begin{proposition}\label{prop2}
Assume that $\vert\vert\vert\rho\vert\vert\vert_\infty <+\infty$ with probability one. Given $q\in\mathbb{N}$, if $\langle\vert\vert\vert\Delta\rho\vert\vert\vert_{\infty}^q\rangle_\rho <+\infty$ and $\langle\vert\vert\vert\nabla\rho\vert\vert\vert_{\infty}^{2q}\rangle_\rho <+\infty$, then for every $t>0$ and $x\in\Lambda$, $\lambda_{q}(x,t)=(q\mu_{x,t})^{-1}$.
\end{proposition}
{\it Proof.} From $\langle\vert\vert\vert\nabla\rho\vert\vert\vert_{\infty}^{2q}\rangle_\rho <+\infty$ and $\langle\vert\vert\vert\Delta\rho\vert\vert\vert_{\infty}^q\rangle_\rho <+\infty$ it follows $\vert\vert\vert\nabla\rho\vert\vert\vert_\infty <+\infty$ and $\vert\vert\vert\Delta\rho\vert\vert\vert_\infty <+\infty$ with probability one. Thus, $\rho$ belongs to the class considered above and Corollary\ \ref{coro1} holds. It remains to prove $\lambda_{q}(x,t)\ge(q\mu_{x,t})^{-1}$.

From ${\cal E}(x,t;s ,\rho)=\Psi(x,t;\eta =i\lambda k(s),\rho)$ it follows that ${\cal E}(x,t;s ,\rho)$ is the solution to\ (\ref{eq3.2}) where the potential is given by\ (\ref{eq3.3}) with the substitution $\varphi_i(x,t)\rightarrow U(x,t;\hat{s})\delta_{1i}$ and $\eta_i\rightarrow i\lambda\| k(s)\|\delta_{1i}=i\lambda\| s\|^2\delta_{1i}$. Equation\ (\ref{eq3.7}) then yields,
\begin{eqnarray}\label{eq4.3}
\vert{\cal E}(x,t;s ,\rho)\vert^q\le&&\left\lbrack A+Bt\left(\vert\vert\vert\Delta\rho\vert\vert\vert_\infty +C\lambda\| s\|^2\right)
+Bt^2\left(\vert\vert\vert\nabla\rho\vert\vert\vert_\infty +D\lambda\| s\|^2\right)^2\right\rbrack^q \nonumber \\
&\times&{\rm e}^{q\lambda\mu_{x,t}\| s\|^2}.
\end{eqnarray}
Let $p_1$ and $p_2$ be two integers such that $0\le p_2\le p_1\le q$. By the Schwartz inequality and the hypotheses, one has,
\begin{eqnarray}\label{eq4.4}
\langle\vert\vert\vert\Delta\rho\vert\vert\vert_\infty^{q-p_1}\vert\vert\vert\nabla\rho\vert\vert\vert_\infty^{2p_2}\rangle_\rho &\le&
\langle\vert\vert\vert\Delta\rho\vert\vert\vert_\infty^{q-p_1+p_2}\rangle_\rho^{\frac{q-p_1}{q-p_1+p_2}}
\langle\vert\vert\vert\nabla\rho\vert\vert\vert_\infty^{2(q-p_1+p_2)}\rangle_\rho^{\frac{p_2}{q-p_1+p_2}}\nonumber \\
&\le&\langle\vert\vert\vert\Delta\rho\vert\vert\vert_\infty^{q}\rangle_\rho^{\frac{q-p_1}{q}}
\langle\vert\vert\vert\nabla\rho\vert\vert\vert_\infty^{2q}\rangle_\rho^{\frac{p_2}{q}}<+\infty,
\end{eqnarray}
\begin{eqnarray}\label{eq4.5}
\langle\vert\vert\vert\Delta\rho\vert\vert\vert_\infty^{q-p_1}\vert\vert\vert\nabla\rho\vert\vert\vert_\infty^{p_2}\rangle_\rho &\le&
\langle\vert\vert\vert\Delta\rho\vert\vert\vert_\infty^{q-p_1+p_2}\rangle_\rho^{\frac{q-p_1}{q-p_1+p_2}}
\langle\vert\vert\vert\nabla\rho\vert\vert\vert_\infty^{q-p_1+p_2}\rangle_\rho^{\frac{p_2}{q-p_1+p_2}}\nonumber \\
&\le&\langle\vert\vert\vert\Delta\rho\vert\vert\vert_\infty^{q}\rangle_\rho^{\frac{q-p_1}{q}}
\langle\vert\vert\vert\nabla\rho\vert\vert\vert_\infty^{2q}\rangle_\rho^{\frac{p_2}{2q}}<+\infty,
\end{eqnarray}
and
\begin{eqnarray}\label{eq4.6}
\langle\vert\vert\vert\nabla\rho\vert\vert\vert_\infty^{2(q-p_1)}\vert\vert\vert\nabla\rho\vert\vert\vert_\infty^{p_2}\rangle_\rho &\le&
\langle\vert\vert\vert\nabla\rho\vert\vert\vert_\infty^{2(q-p_1+p_2)}\rangle_\rho^{\frac{q-p_1}{q-p_1+p_2}}
\langle\vert\vert\vert\nabla\rho\vert\vert\vert_\infty^{q-p_1+p_2}\rangle_\rho^{\frac{p_2}{q-p_1+p_2}}\nonumber \\
&\le&\langle\vert\vert\vert\nabla\rho\vert\vert\vert_\infty^{2q}\rangle_\rho^{\frac{2(q-p_1)+p_2}{2q}}<+\infty.
\end{eqnarray}
From\ (\ref{eq4.4})-(\ref{eq4.6}) it follows that there exists a polynomial of degree $2q$, $P_{2q}(\cdot)$, such that $\langle\vert{\cal E}(x,t;s ,\rho)\vert^q\rangle_\rho$ is bounded by
$$
\langle\vert{\cal E}(x,t;s ,\rho)\vert^q\rangle_\rho\le P_{2q}(\| s\|^2){\rm e}^{q\lambda\mu_{x,t}\| s\|^2}.
$$
Thus, for every $\lambda <(q\mu_{x,t})^{-1}$ and almost every realization of $\rho$,
\begin{equation}\label{eq4.7}
\langle\vert {\cal E}(x,t)\vert^q\rangle_{\rho ,s}\le
\int\cdots\int_{{\mathbb C}^M} P_{2q}(\| s\|^2){\rm e}^{(q\lambda\mu_{x,t}-1)\| s\|^2}
\prod_{n=1}^M \frac{d^2s_n}{\pi}<+\infty ,
\end{equation}
which proves $\lambda_{q}(x,t)\ge (q\mu_{x,t})^{-1}$. $\square$
%
%
\section{Summary and perspectives}\label{sec5}
In this paper, we have studied the divergence of the solution to a Schr\"odinger-type amplifier driven by the square of a Gaussian noise in presence of a random potential. For restricted but quite wide classes of driver and potential , we have explicitely determined the values of the coupling constant at which the moments of the solution diverge. Both moments w.r.t. the driver for almost every realization of the potential, and moments w.r.t. both the driver and the potential have been considered (with a slightly reduced class of potential in the latter case). We have followed the same approach as in\ \cite{MCL} in terms of distributional formulation of the solution to\ (\ref{eq1.1}) and use of the Paley-Wiener theorem.

Our results show that the divergence is not affected by the random potential, in the sense that it occurs at exactly the same coupling constant as what was found in\ \cite{MCL} without a potential. It follows {\it a fortiori} that the breakdown of the amplifier is not affected by the possible existence of a localized regime in the amplification free limit.

As far as we know, there is no general simple criterium to decide whether a potential belongs to the class(es) considered here. Nevertheless, sufficient conditions can be given when the potential is a zero mean homogeneous and stationary Gaussian field. In this case it can be shown\ \cite{Adl} that Propositions\ \ref{prop1} and\ \ref{prop2} hold if the correlation function of the potential is $C^6$ in space and $C^2$ in time. It follows in particular that our results apply in all the cases where the potential has a smooth ($C^\infty$) correlation function. In laser-plasma interaction, our results are expected to hold provided that the underlying hydrodynamic evolution can insure smoothness of the plasma density at the scale of the coarse grained description\ (\ref{eq1.1}).

We note that our conclusions carry over to cases where $S$ and $\rho$ may be correlated, as they are in laser-plasma interaction. For instance, consider a potential $\rho = \rho_0+\rho_1$ where $\rho_0$ is independent of $S$ and $\rho_1\propto\vert S\vert^2$ is the $S$-induced potential perturbation. Putting this potential on the right-hand side of\ (\ref{eq1.1}) amounts to add a non zero imaginary part to $\lambda$. Now, the Paley-Wiener theorem involves the real part of $\lambda$ only, not its imaginary part. Consequently, the whole machinery of the sections III and IV of\ \cite{MCL} is unaffected by the presence of $\rho_1$, as is the value of the critical coupling.

In most applications, space average is expected to give a more appropriate description of the measured amplification than ensemble average. For a finite system, space average is almost surely finite and it seems difficult to find an unambiguous definition of the critical coupling in that case. In order to get clear-cut results one must let the system size go to infinity. Although the random potential does not affect the value of the critical coupling, it may affect the speed at which space average diverges above the critical coupling as the system size goes to infinity. From a practical point of view, it would be interesting to find out whether this effect does exist and, if so, to investigate it. Such a study, which will presumably require the use of numerical simulations of\ (\ref{eq1.1}), will be the subject of a future work.
\section*{Acknowledgements}
Ph. M. thanks Harvey A. Rose for providing many valuable insights. We also thank Joel L. Lebowitz for useful discussions.
%
%
\appendix*
\section{Determination of the support of $\bm{g_{(x,T)}^{(m)}}$}
Let $\hat{s}\equiv s/\| s\|$ be the direction of a given $s$ in $\mathbb{C}^M$ and $\hat{k}(s)= k(s)/\| k(s)\|$ the corresponding direction in $\mathbb{R}^N$, with $k(s)$ defined above Equation\ (\ref{eq3.1}). Let $g_{x,t;\rho}^{(m)}(u)$ be a distribution with compact support on ${\mathbb R}$ whose Fourier transform, $\Psi^{(m)}(x,t;\eta ,\rho)\equiv({\cal F}g_{x,t;\rho}^{(m)})(\eta)$ with $\eta\in {\mathbb R}$, is the solution to\ (\ref{eq3.2}) with $V(x,t;\eta)=\eta U(x,t;\hat{s})$, where $U(x,t;\hat{s})=\sum_{i=1}^N\hat{k}(s)_i \varphi_i(x,t)$. This appendix is devoted to the determination of the support of $g_{(x,t;\rho)}^{(m)}$. We have modified the notation used in the text to make the dependence on $m$ explicit.
\paragraph*{}We begin with a technical lemma that will be useful in the sequel. Let $C_0^{\infty}({\mathbb R})$ denote the set of all smooth compactly supported functions in ${\mathbb R}$, and ${\mathbb C}^+ \equiv\lbrace m\in {\mathbb C}:\ {\rm Im}(m)>0\rbrace$.
\begin{lemma}\label{lem3}
For every $t>0$, $x\in\Lambda$, $f\in C_0^{\infty}({\mathbb R})$, $z\in\mathbb{C}$ (resp. $m\in\mathbb{C}^+$), and almost every realization of $\rho$, $\int_{{\mathbb R}} g_{x,t;z\rho}^{(m)}(u)f(u)\, du$ is an analytic function of $m\in\mathbb{C}^+$ (resp. $z\in\mathbb{C}$), and $\int_{{\mathbb R}} g_{x,t;z\rho}^{(m)}(u)f(u)\, du=\lim_{\gamma\rightarrow 0^+} \int_{{\mathbb R}} g_{x,t;z\rho}^{(m+i\gamma)}(u)f(u)\, du$ for each real $m\ne 0$ and every $z\in\mathbb{C}$.
\end{lemma}
{\it Proof.} First we prove that, for almost every realization of $\rho$, (i) $\Psi^{(m)}(\cdot ,t;\eta ,z\rho)\in L^2(\Lambda)$ is analytic in $(m,\eta,z)\in {\mathbb C}^+\times{\mathbb C}^2$; and (ii) $\forall\, (\eta ,z)\in {\mathbb C}^2$, $\Psi^{(m)}(\cdot ,t;\eta ,z\rho)=\lim_{\gamma\rightarrow 0^+}\Psi^{(m+i\gamma)}(\cdot ,t;\eta ,z\rho)$ in $L^2(\Lambda)$ for each real $m\ne 0$\ \cite{Yajima}. Define
$$
{\cal U}_m(t)=\exp\left(\frac{it\Delta}{2m}\right),\ \ t\ge 0,\ \ m\in{\mathbb C}^+,
$$
and write the initial value problem\ (\ref{eq3.2}) in the the form of integral equation
$$
\Psi^{(m)}(t)=1-i\int_0^t {\cal U}_m(t-\tau)V(\tau)\Psi^{(m)}(\tau)\, d\tau .
$$
Here $V(t)$ is the multiplication operator with $z\rho(x,t)+\eta U(x,t;\hat{s})$. Let ${\cal B}$ denote the space of bounded operators in $L^2(\Lambda)$. By Fourier series expansion, it is evident that (a) $\vert\vert {\cal U}_m(t)\Psi^{(m)}(t)\vert\vert_2\le\vert\vert\Psi^{(m)}(t)\vert\vert_2$, viz. ${\cal U}_m(t)\in{\cal B}$ and $\vert\vert {\cal U}_m(t)\vert\vert\le 1$; (b) the function $\lbrack 0,\infty)\times (\overline{{\mathbb C}^+}\backslash\lbrace 0\rbrace)\ni (t,m)\rightarrow {\cal U}_m(t)\in{\cal B}$ is strongly continuous [viz. $(t,m)\rightarrow {\cal U}_m(t)f\in L^2(\Lambda)$ is continuous for every $f\in L^2(\Lambda)$]; and (c) for every $t\ge 0$, $m\rightarrow {\cal U}_m(t)\in{\cal B}$ is analytic for $m\in{\mathbb C}^+$ and $(d/dm){\cal U}_m(t)$ is norm continuous w.r.t. $(t,m)\in\lbrack 0,\infty)\times{\mathbb C}^+$. It follows from the boundedness of $U$ and $\rho$ (almost surely) that the Dyson expansion\ \cite{RS}
\begin{eqnarray*}
&&D_m(t)={\cal U}_m(t)-i\int_0^t{\cal U}_m(t-\tau)V(\tau){\cal U}_m(\tau)\, d\tau
+\cdots + \nonumber \\
&&(-i)^n\int_{0<\tau_1<\cdots <\tau_n<t}{\cal U}_m(t-\tau_n)V(\tau_n)
\cdots V(\tau_1){\cal U}_m(\tau_1)\, d\tau_1\cdots d\tau_n +\cdots
\end{eqnarray*}
converges in the operator norm of ${\cal B}$ uniformly w.r.t. $(t,m)$ in every compact subset of $\lbrack 0,\infty)\times (\overline{{\mathbb C}^+}\backslash\lbrace 0\rbrace)$. Thus, the operator $D_m(t)$ enjoys the same properties (b) and (c) mentioned above as an operator valued function of $t$ and $m$. It is easy to check that $D_m(t)$ defines the propagator for\ (\ref{eq3.2}) and is unitary if $m$, $\eta$, and $z$ are real. Hence the solution to\ (\ref{eq3.2}) satisfies the properties (i) and (ii). From now on $t>0$ is fixed.

Without loss of generality, we take for $\Lambda$ the $d$-dimensional torus of length unity. For every $n\in{\mathbb Z}^d$, define $\hat{\Psi}_n^{(m)}(t;\eta ,z\rho)=\int_{\Lambda}\Psi^{(m)}(x ,t;\eta ,z\rho)\exp(-2i\pi n\cdot x)\, d^dx$. From this expression, the compactness of $\Lambda$, and the Schwartz inequality, it is easily seen that properties (i) and (ii) imply that for every $n\in{\mathbb Z}^d$ and almost every realization of $\rho$, (iii) $\hat{\Psi}_n^{(m)}(t;\eta ,z\rho)$ is analytic in $(m,\eta,z)\in {\mathbb C}^+\times{\mathbb C}^2$; and (iv) $\forall\, (\eta ,z)\in {\mathbb C}^2$, $\hat{\Psi}_n^{(m)}(t;\eta ,z\rho)=\lim_{\gamma\rightarrow 0^+}\hat{\Psi}_n^{(m+i\gamma)}(t;\eta ,z\rho)$ for each real $m\ne 0$.

Now, $\Psi^{(m)}(\cdot ,t;\eta ,z\rho)$ is actually in $H^2(\Lambda)$\ \cite{note3}. As a result, $\Psi^{(m)}(x ,t;\eta ,z\rho)$ is a continuous function of $x\in\Lambda$, and by bounding its $H^2(\Lambda)$-norm in the same way as in \ \cite{MCL} it is not difficult to prove that, for $d\le 3$, $\sum_{\| n\|\le R}\hat{\Psi}_n^{(m)}(t;\eta ,z\rho)\exp(2i\pi n\cdot x)$ converges to $\Psi^{(m)}(x ,t;\eta ,z\rho)$ uniformly w.r.t. $(x,m)$ in $\Lambda\times (\overline{{\mathbb C}^+}\backslash\lbrace 0\rbrace)$ and $(\eta ,z)$ in every compact subset of ${\mathbb C}^2$ as $R\rightarrow +\infty$. This result together with (iii), (iv), and Morera's theorem\ \cite{Con} imply that for every $x\in\Lambda$ and almost every realization of $\rho$, (v) $\Psi^{(m)}(x,t;\eta ,z\rho)$ is analytic in $(m,\eta,z)\in {\mathbb C}^+\times{\mathbb C}^2$; and (vi) $\forall\, (\eta ,z)\in {\mathbb C}^2$, $\Psi^{(m)}(x,t;\eta ,z\rho)=\lim_{\gamma\rightarrow 0^+}\Psi^{(m+i\gamma)}(x,t;\eta ,z\rho)$ for each real $m\ne 0$.

Fix $x\in\Lambda$. The bound\ (\ref{eq3.7}) has been obtained for $\eta$ complex and $z$ real ($z=1$). One might as well take $\eta$ real and $z$ complex. In that case, $\vert\Psi^{(m)}(x,t;\eta ,z\rho)\vert$ is bounded by an expression similar to\ (\ref{eq3.7}) with $\vert z\vert$ instead of $\vert\eta\vert$ in the exponential. It follows that, if $\eta$ is real, then for every $m$ in $\overline{{\mathbb C}^+}\backslash\lbrace 0\rbrace$ and $z$ in a compact subset of ${\mathbb C}$ one can bound $\vert\Psi^{(m)}(x,t;\eta ,z\rho)\vert$ with a polynomial of $\vert\eta\vert$ the coefficients of which are independent of $m$ and $z$. As a result, $\vert\Psi^{(m)}(x,t;\eta ,z\rho)(\mathcal{F}f)(-\eta)\vert$ is bounded by an integrable function of $\eta$ independent of $m$ and $z$. Thus, by dominated convergence, properties (v) and (vi) imply that
$$
\int_{{\mathbb R}} g_{x,t;z\rho}^{(m)}(u)f(u)\, du\equiv\int_{{\mathbb R}}\Psi^{(m)}(x,t;\eta ,z\rho)(\mathcal{F}f)(-\eta)\frac{d\eta}{2\pi}
$$
is a continuous function of $m$ in $\overline{{\mathbb C}^+}\backslash\lbrace 0\rbrace$ and $z$ in every compact subset of ${\mathbb C}$. The continuity in $m$ proves the second part of Lemma\ \ref{lem3}. The first part can be proved straightforwardly as an application of Morera's theorem\ \cite{Con}. The line of the proof is as follows: (a) integrate both sides of the above identity w.r.t. $m$ (or $z$) along any closed path in ${\mathbb C}^+$ (or ${\mathbb C}$); (b) by Fubini's theorem the $m$- (or $z$-) and $\eta$-integrals can be interchanged and the result follows immediately from (v), Cauchy's theorem, and Morera's theorem. $\square$

Let $a=\inf_{x(\cdot)\in B(x,t)}\int_0^tU(x(\tau),\tau;\hat{s})\, d\tau$ and $b=\sup_{x(\cdot)\in B(x,t)}\int_0^tU(x(\tau),\tau;\hat{s})\, d\tau =H_{x,t}(\hat{s})$. One has the following Lemma:
\begin{lemma}\label{lem4}
For every $t>0$, $x\in\Lambda$, $m\in\overline{{\mathbb C}^+}\backslash\lbrace 0\rbrace$, and almost every realization of $\rho$, the support of $g_{x,t;\rho}^{(m)}$ is equal to $\lbrack a,b\rbrack$.
\end{lemma}
{\it Proof.} First, consider the case $m=i\gamma$ and $z=-iy$, with $\gamma >0$ and $y>0$. Denote by $\alpha\lbrack x(\cdot)\rbrack$ the functional $\alpha\lbrack x(\cdot)\rbrack\equiv\int_0^tU(x(\tau),\tau;\hat{s})\, d\tau$. It is proved in\ \cite{MCL}, Appendix B, that $\alpha\lbrack x(\cdot)\rbrack$ is a continuous functional of $x(\cdot)\in B(x,t)$ with the uniform norm on $\lbrack 0,t\rbrack$. Let $h\in C_0^{\infty}({\mathbb R})$ a real positive test function with support in $\lbrack a,b\rbrack$ and $\sup_{u\in {\mathbb R}}h(u)=1$. From the continuity of $\alpha\lbrack x(\cdot)\rbrack$ it follows that $\exists x_0(\cdot)\in B(x,t)$ such that $h(\alpha\lbrack x_0(\cdot)\rbrack)=1$. By continuity of $h$ and $\alpha\lbrack x(\cdot)\rbrack$ it follows that $\forall\varepsilon >0$, $\exists\delta >0$ such that $\vert h(\alpha\lbrack x(\cdot)\rbrack)-1\vert <\varepsilon$ for every $x(\cdot)\in B_0(\delta)\equiv\lbrace x(\cdot)\in B(x,t):\ \sup_{0\le\tau\le t}\| x(\tau)-x_0(\tau)\| <\delta\rbrace$. Take $\varepsilon =1/2$, in this case $h(\alpha\lbrack x(\cdot)\rbrack)>1/2$ for every $x(\cdot)\in B_0(\delta)$ and for every realization of $\rho$ one has
\begin{eqnarray*}
\int_{{\mathbb R}} g_{x,t;-iy\rho}^{(i\gamma)}(u)h(u)\, du &=&
\int_{x(\cdot)\in B(x,t)}{\rm e}^{
-\int_0^t\left\lbrack\frac{\gamma}{2}\dot{x}(\tau)^2+y\rho(x(\tau),\tau)\right\rbrack\, d\tau}
h(\alpha\lbrack x(\cdot)\rbrack)\,
d\lbrack x(\cdot)\rbrack \nonumber \\
&\ge&\int_{x(\cdot)\in B_0(\delta)}
{\rm e}^{
-\int_0^t\left\lbrack\frac{\gamma}{2}\dot{x}(\tau)^2+y\rho(x(\tau),\tau)\right\rbrack\, d\tau}
h(\alpha\lbrack x(\cdot)\rbrack)\,
d\lbrack x(\cdot)\rbrack \\
&>&\frac{1}{2}\int_{x(\cdot)\in B_0(\delta)}
{\rm e}^{-\int_0^t\left\lbrack\frac{\gamma}{2}\dot{x}(\tau)^2+y\rho(x(\tau),\tau)\right\rbrack\, d\tau}
d\lbrack x(\cdot)\rbrack ,
\end{eqnarray*}
which is bounded below by
$$
\int_{{\mathbb R}} g_{x,t;-iy\rho}^{(i\gamma)}(u)h(u)\, du>\frac{1}{2}{\rm e}^{-yt\vert\vert\vert\rho\vert\vert\vert_\infty}
\int_{x(\cdot)\in B_0(\delta)}
{\rm e}^{-\frac{\gamma}{2}\int_0^t\dot{x}(\tau)^2 d\tau}
d\lbrack x(\cdot)\rbrack .
$$
Since the set of the Brownian paths $x(\cdot)$ that are in $B_0(\delta)$ has a strictly positive Wiener measure and $\vert\vert\vert\rho\vert\vert\vert_\infty <+\infty$ almost surely, the last term is (almost surely) strictly positive and one finds, for almost every realization of $\rho$,
\begin{equation}\label{eqb1}
\int_{{\mathbb R}} g_{x,t;-iy\rho}^{(i\gamma)}(u)h(u)\, du >0.
\end{equation}
If there was an open subset of $\lbrack a,b\rbrack$ not intersecting the support of $g_{x,t;-iy\rho}^{(i\gamma)}$, it would be possible to choose the support of $h$ outside the one of $g_{x,t;-iy\rho}^{(i\gamma)}$, yielding $\int g_{x,t;-iy\rho}^{(i\gamma)}(u)h(u)\, du=0$ in contradiction with Eq.\ (\ref{eqb1}). Thus, for every $x\in\Lambda$, $\gamma >0$, $y>0$, and almost every realization of $\rho$, the support of $g_{x,t;-iy\rho}^{(i\gamma)}$ is equal to $\lbrack a,b\rbrack$.

Consider now the case $m=i\gamma$ ,$\gamma >0$, and $z\in\mathbb{C}$. Fix a realization of $\rho$ for which both Lemma\ \ref{lem3} and\ (\ref{eqb1}) hold, and assume that there is an open subset of $\lbrack a,b\rbrack$ not intersecting the support of $g_{x,t;z\rho}^{(i\gamma)}$. In this case it is possible to choose the support of $h$ outside the one of $g_{x,t;z\rho}^{(i\gamma)}$, yielding $\int g_{x,t;z\rho}^{(i\gamma)}(u)h(u)\, du =0$. By Lemma\ \ref{lem3} the support of $g_{x,t;z\rho}^{(i\gamma)}$ must vary continuously with $z\in\mathbb{C}$, whence the support of $h$ can be taken small enough such that there is a open subset ${\cal V}(z)\subset\mathbb{C}$ with $z\in {\cal V}(z)$ and $\int g_{x,t;z\rho}^{(i\gamma)}(u)h(u)\, du=0$ identically in ${\cal V}(z)$. From the analyticity of $\int g_{x,t;z\rho}^{(i\gamma)}(u)h(u)\, du$ in $z$ on $\mathbb{C}$ (Lemma\ \ref{lem3}), it follows immediately that $\int g_{x,t;z\rho}^{(i\gamma)}(u)h(u)\, du=0$ identically in all $\mathbb{C}$, in contradiction with Eq.\ (\ref{eqb1}). Since the set of the realizations of $\rho$ that do not fulfill Lemma\ \ref{lem3} or\ (\ref{eqb1}) is of zero probability, one finds that for every $x\in\Lambda$, $\gamma >0$, $z\in\mathbb{C}$, and almost every realization of $\rho$, the support of $g_{x,t;z\rho}^{(i\gamma)}$ is equal to $\lbrack a,b\rbrack$.

Consider finally the case $m\in\overline{{\mathbb C}^+}\backslash\lbrace 0\rbrace$ and $z=1$. Again, fix a realization of $\rho$ for which both Lemma\ \ref{lem3} and\ (\ref{eqb1}) hold, and assume that there is an open subset of $\lbrack a,b\rbrack$ not intersecting the support of $g_{x,t;\rho}^{(m)}$. Since the support of $g_{x,t;\rho}^{(i\gamma)}$ ($\gamma >0$) is equal to $\lbrack a,b\rbrack$ (see above), it is always possible to choose a test function $f\in C_0^{\infty}({\mathbb R})$ the support of which lies outside the one of $g_{x,t;\rho}^{(m)}$, yielding $\int g_{x,t;\rho}^{(m)}(u)f(u)\, du =0$, and such that
\begin{equation}\label{eqb2}
\int_{{\mathbb R}} g_{x,t;\rho}^{(i\gamma)}(u)f(u)\, du >0.
\end{equation}
By Lemma\ \ref{lem3} the support of $g_{x,t;\rho}^{(m)}$ must vary continuously with $m\in\overline{{\mathbb C}^+}$, whence the support of $f$ can be taken small enough such that there is a open subset ${\cal V}(m)\subset {\mathbb C}^+$ with $m\in\overline{{\cal V}(m)}$ and $\int g_{x,t;\rho}^{(m)}(u)f(u)\, du=0$ identically in ${\cal V}(m)$. From the analyticity of $\int g_{x,t;\rho}^{(m)}(u)f(u)\, du$ in $m$ on ${\mathbb C}^+$ (Lemma\ \ref{lem3}), it follows immediately that $\int g_{x,t\rho}^{(m)}(u)f(u)\, du=0$ identically in all ${\mathbb C}^+$, in contradiction with Eq.\ (\ref{eqb2}). The fact that the set of the realizations of $\rho$ that do not fulfill Lemma\ \ref{lem3} or\ (\ref{eqb1}) is of zero probability completes the proof of Lemma\ \ref{lem4}. $\square$

It follows from Lemma\ \ref{lem4} that $\sup\lbrace v\in\mathbb{R}:v\in {\rm supp}g_{x,t;\rho}^{(m)}\rbrace =b=H_{x,t}(\hat{s})$ for almost every realization of $\rho$, which is the result used in the proof of Lemma\ \ref{lem2}.
%
%

%
%

\begin{thebibliography}{1}
\bibitem{note2} The proof is for $\eta\in{\mathbb C}$. The generalization to $\eta\in{\mathbb C}^N$ is straightforward.
\bibitem{note3} This is ensured by the fact that the $L^2(\Lambda)$-norms of both $\Psi$ and $\Delta\Psi$ exist (see\ \cite{MCL}).
\end{thebibliography}
\end{document}